\documentclass[runningheads]{llncs}
\usepackage{graphicx}
\begin{document}
\title{Real-time 3D Light-field Viewing with Eye-tracking on Conventional Displays\thanks{This work was supported by JSPS KAKENHI under Grant 24K20797.}}
\titlerunning{3D Light-field Viewing on Conventional Displays}
%
\author{Trung Hieu PHAM\inst{1} \and
Chanh Minh TRAN\inst{1} \and
Eiji KAMIOKA\inst{1} \and
PHAN Xuan Tan\inst{1}\thanks{Corresponding author: \email{tanpx@shibaura-it.ac.jp}}}
\authorrunning{T.H. Pham et al.}
%
\institute{Shibaura Institute of Technology, Tokyo, Japan\\
\email{\{ma24503, tran.chanh.r4, kamioka, tanpx\}@shibaura-it.ac.jp}
}
\maketitle              
\begin{abstract}
Creating immersive 3D visual experiences typically requires expensive and specialized hardware such as VR headsets, autostereoscopic displays, or active shutter glasses. These constraints limit the accessibility and everyday use of 3D visualization technologies in resource-constrained settings.
To address this, we propose a low-cost system that enables real-time 3D light-field viewing using only a standard 2D monitor, a conventional RGB webcam, and red-cyan anaglyph glasses. The system integrates real-time eye-tracking to dynamically adapt the displayed light-field image to the user's head position with a lightweight rendering pipeline that selects and composites stereoscopic views from pre-captured light-field data. The resulting anaglyph image is updated in real-time, creating a more immersive and responsive 3D experience.
The system operates entirely on CPU and maintains a stable frame rate of 30 FPS, confirming its feasibility on typical consumer-grade hardware. All of these highlight the potential of our approach as an accessible platform for interactive 3D applications in education, digital media, and beyond.

\end{abstract}
\section{Introduction}\label{sec1}
The ability to perceive and interact with three-dimensional (3D) imagery has significantly impacted a wide range of domains, from entertainment~\cite{hattler2023expanded,pierce1997image} to education~\cite{wang20243d}, and medical imaging~\cite{robb1989interactive,johnsonchris2022review}. Over the past several decades, advancements in display~\cite{yang2016see} and interaction technologies have brought immersive 3D experiences closer to mainstream use. These experiences typically rely on specialized hardware, such as head-mounted displays (HMDs) for virtual reality (VR), augmented reality (AR) headsets, or stereoscopic monitors using polarized or active-shutter glasses, to deliver distinct images to each eye and create a convincing illusion of depth. While such systems provide rich spatial cues and user immersion, they often entail high costs, intricate setup procedures, and ergonomic constraints, which limit their viability for everyday or large-scale deployment in informal environments such as homes or classrooms.

To bridge the gap between high-quality immersive visualization and practical accessibility, simpler stereoscopic approaches~\cite{cosmo2014low,powell2016getting} have emerged. Among these, anaglyph glasses, which use colored filters to deliver different images to each eye, stand out due to their extremely low cost and compatibility with standard display technology~\cite{deakyne2021development}. By requiring only an ordinary monitor and inexpensive paper or plastic glasses, anaglyph techniques allow basic 3D effects to be experienced without the need for specialized or proprietary hardware. However, this simplicity comes with significant trade-offs. Anaglyph rendering has long been associated with limited color fidelity, ghosting~\cite{woods2004ghosting}, and a lack of dynamic interaction capabilities, such as head-coupled parallax, which impairs the realism of the 3D experience and reduces viewer comfort during extended use.

Recent advances in real-time computer vision~\cite{ryan2021real,adhanom2023eye,fang2022alphapose}, particularly in the domains of facial and eye-tracking using ordinary RGB cameras, open new possibilities for enhancing anaglyph-based systems. By leveraging these technologies, it becomes feasible to introduce dynamic perspective adjustment based on user position, restoring an essential depth cue known as motion parallax. When combined with pre-rendered or captured light-field imagery, this approach enables natural and immersive 3D experiences, even on basic display hardware.

Light-field imaging~\cite{10.1145/237170.237199,10.1145/237170.237200} plays a central role in this work as a powerful method for capturing and rendering 3D content. Light-field imaging is an innovative technology that revolutionizes the process of capturing and perceiving images. It captures not only the intensity and color of light rays at different points in space but also their directional information. This enables the generation of multiple views from different angles, which is crucial for achieving viewpoint-dependent rendering and dynamic parallax. Furthermore, light-field content can be acquired easily by using many types of light-field cameras on the market. It alleviates the burden on creating a 3D content, making it easier for normal end-user.

In this paper, we introduce a low-cost, eye-tracked light-field viewing system that provides interactive 3D visualization capabilities using standard 2D monitors and inexpensive anaglyph glasses. Our system relies on a single ordinary RGB camera to track the viewer's eye position in real-time and adjust the displayed content to match the user's viewpoint. Light-field image data is reprojected accordingly to provide coherent parallax and stereo cues that respond to head motion, resulting in a more immersive and convincing sense of depth. Notably, this system requires no special display panels, headsets, or high-end sensors, making it broadly accessible and easy to deploy.

The proposed system integrates two key components: (1) robust, low-latency eye-tracking and real-time head-coupled perspective adjustment using computer vision techniques; and (2) a rendering pipeline that selects appropriate views from a light-field scene, optimized for anaglyph stereo display. Together, these elements form a practical solution for light-field content exploration, with potential applications in education, 3D data analysis, digital art, and beyond.

In this paper, we have several contributions. First, we present the design and implementation of a functional prototype that achieves real-time performance on consumer-grade hardware using easily accessible components. Next, we demonstrate that our system provides natural perceived depth and parallax that reflect real-world perceptual effects. Finally, our system does not require specialized hardware, making it suitable for a broad deployment in settings such as classrooms, laboratories, and home environments.

The rest of this paper is organized as follows. Section~\ref{sec: related_work} reviews related work in 3D visualization technologies, light-field imaging and eye-tracking with Mediapipe~\cite{lugaresi2019mediapipe}. Section~\ref{sec: proposed framework} describes our system architecture, including hardware components, tracking algorithms, and the rendering pipeline. Section~\ref{sec: experiment} presents implementation details, and experimental results. Section~\ref{sec: discussion} discusses system limitations, potential applications, and directions for future work. Section~\ref{sec: conclusion} concludes with a summary of our contributions and their implications for accessible 3D content interaction.

\section{Related Work}\label{sec: related_work}

\subsection{3D Visualization Technologies}

The development of 3D visualization technologies has produced a rich spectrum of systems that differ in terms of immersion, accessibility, and visual fidelity. These systems can generally be categorized into immersive head-mounted, glasses-free (autostereoscopic), and glasses-based stereoscopic solutions~\cite{gao2016latest,geng2013three}.

HMDs used in VR and AR provide fully immersive 3D experiences by enclosing the user’s field of view and tracking head movement. Devices like the Meta Quest 3~\cite{metaquest2024}, HTC Vive~\cite{htcvive}, and Microsoft HoloLens~\cite{hololens} enable rich interactivity and spatial immersion by combining stereoscopic rendering with six degrees of freedom (6DoF) motion tracking. These systems provide excellent depth cues including motion parallax, and stereo disparity, but are often constrained by high cost, user discomfort, and content development overhead~\cite{martirosov2017cyber,kim2021user}.

Autostereoscopic displays or glasses-free displays offer 3D visualization without requiring users to wear glasses. These systems include parallax barrier displays, lenticular lens screens, and more advanced volumetric and light-field displays. Notable examples include the Nintendo 3DS~\cite{nintendo3ds}, which uses a parallax barrier to present distinct images to each eye, and the Looking Glass series~\cite{lookingglass} of volumetric displays, which presents multiple views of a 3D scene simultaneously to support motion parallax and accommodate multiple viewers. However, Looking Glass displays are limited in both image quality and viewing angle due to the need to render many views at once. These systems typically work best if you are standing in the right position, and may lose the 3D effect if you move too far off to the side~\cite{cserkaszky2018light}. In contrast, Sony's Eye-Sensing Light Field Display~\cite{aoyama202148} employs lenticular lens with real-time eye tracking to deliver a high-resolution 3D experience by rendering views only for the user's two eyes. This enables wider viewing angles and superior image quality, though it restricts the display to single-person use. Lenticular lens screens work like those 3D postcards or stickers. When you move your head, the image appears to shift. This happens because each tiny lens bends light in a way that sends a different image to each of your eyes, creating a 3D effect. These technologies allow for more natural interaction and viewing comfort. However, all of these displays require specialized and expensive displays, making them difficult for everyone to access.

Another type of display that can provide the 3D viewing experience is stereoscopic display that requires glasses. These displays have been more prevalent in consumer and professional applications due to their relative simplicity and effectiveness. Polarized glasses are widely used in cinema and some 3D televisions, employing filters to separate left and right images projected onto a compatible screen~\cite{richtberg2017use}. Active shutter glasses alternate opaque lenses in sync with the display’s refresh rate, delivering high-quality stereoscopic images at the cost of increased hardware complexity. Anaglyph glasses, which use color filters to separate the images, represent the most affordable and accessible option~\cite{peddie2013history}. Although they suffer from color distortion and limited depth accuracy, their compatibility with standard displays and low price have made them easy to be widely accessed.

\subsection{Light-field Imaging}

Light-field imaging~\cite{10.1145/237170.237199,10.1145/237170.237200} has emerged as a powerful method for capturing and representing 3D content by recording the direction and intensity of light rays from a scene. Unlike traditional 2D photographs, which capture a flat image from a single viewpoint, light-field images preserve the spatial structure of a scene, enabling advanced features such as refocusing, viewpoint interpolation, and depth estimation. This makes light-field content particularly suitable for stereoscopic or autostereoscopic displays, where dynamic view rendering is essential for delivering convincing depth perception.

From a practical standpoint, light-field acquisition has become increasingly user-friendly due to advances in both hardware and software. Commercial light-field cameras like the Lytro Illum~\cite{lytroillum} and Raytrix R-series~\cite{raytrixR} have popularized microlens array technology, where a dense array of small lenses is placed in front of the image sensor to capture multiple perspectives of the scene. This compact and self-contained design makes light-field imaging feasible for photographers, educators, and researchers without the need for complex setups. Meanwhile, camera array systems such as those used in the Stanford Light Field Archive~\cite{10.1145/1186822.1073259,10.1145/237170.237199} dataset employ multiple synchronized cameras arranged in a grid or arc to achieve dense light-field acquisition. These systems can capture high-resolution light-fields with greater angular diversity, supporting more realistic view synthesis and spatial effects. Though more expensive and space-consuming, camera arrays are ideal for content creators seeking high-fidelity data for visual effects, virtual reality, or scientific applications. There are also mobile-scale solutions like the Pelican Imaging array camera~\cite{10.1145/2508363.2508390}, which integrate light-field capture capabilities into a small sensor array, ideal for portable devices. These have demonstrated how light-field capabilities can be embedded into smartphones and tablets. Combined with computational photography techniques and powerful processing units, this opens the door for casual users to produce 3D-capable content using everyday devices.

These technologies empower users to create their own 3D content with relative ease. This ease of content creation makes light-field imaging a compelling foundation for 3D visualization systems that aim to be widely deployable and adaptable to conventional hardware.

\subsection{Eye-tracking with MediaPipe}

Conventional eye-tracking methods typically rely on specialized hardware such as infrared (IR) cameras~\cite{coetzer2014development,guestrin2006general} or wearable devices~\cite{VIDAL20121306}. While these systems provide high accuracy and robustness, they come with significant limitations, including high cost, restricted portability, and reliance on external calibration.

In contrast, MediaPipe~\cite{lugaresi2019mediapipe}, an open-source framework developed by Google, offers a lightweight, real-time alternative that is particularly suitable for applications requiring responsive eye-tracking without dedicated sensors. With the assumption that the user always look at the object on the monitor, eye-tracking can be achieved through detecting iris of the user. MediaPipe’s Face Mesh model~\cite{kartynnik2019real} detects 468 facial landmarks, including a dense set of points around the eye region, enabling reliable estimation of eye position using only an RGB camera. The framework is optimized for speed, achieving real-time inference even on mobile or embedded platforms, and it leverages efficient GPU acceleration when available. Unlike many deep learning models that require heavy computational resources, MediaPipe is designed with cross-platform performance in mind, allowing seamless deployment across desktops, mobile devices, and embedded systems.

These attributes make MediaPipe especially appropriate for systems like ours, where low-latency eye position detection is crucial for dynamically adjusting the rendered view of light-field images. Its minimal setup and calibration-free operation reduce deployment complexity and make it accessible for broader applications in real-world environments.

\section{Proposed Framework}\label{sec: proposed framework}
\subsection{Overview}

\begin{figure}
   \includegraphics[width = \textwidth]{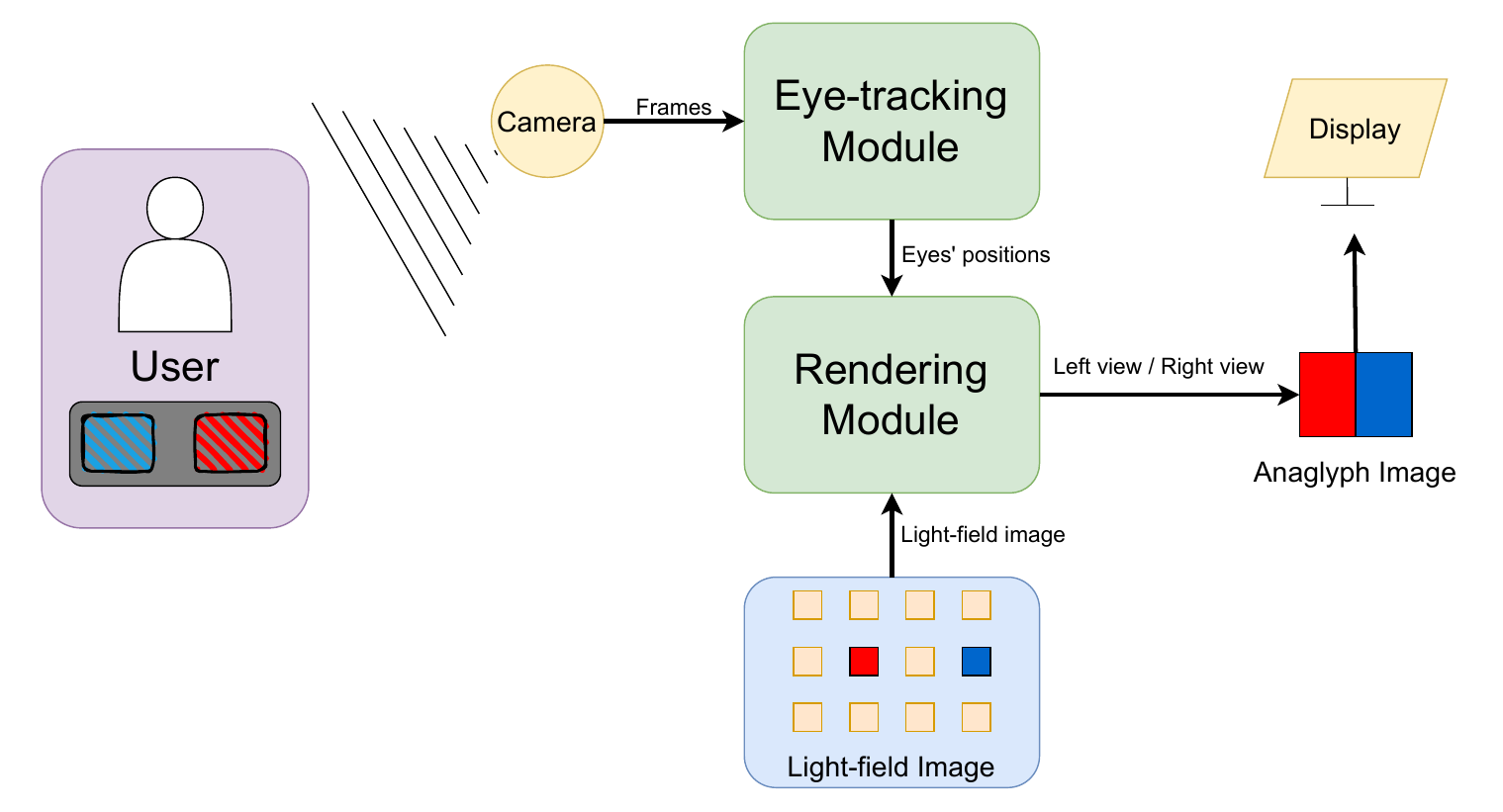}
   \caption{The proposed system has an Eye-tracking Module to get the positions of eyes and a Rendering Module to render the corresponding view to display.}
   \label{fig:overview}
\end{figure}

This research presents a novel approach to 3D visualization by combining standard hardware components with real-time eye-tracking and light-field rendering as demonstrated in Figure~\ref{fig:overview}. The system architecture consists of two primary components: an eye-tracking module and an anaglyph light-field rendering module, all integrated into a real-time application.

The hardware setup is intentionally minimalistic, requiring only components that are widely available and affordable. The system employs a conventional webcam positioned above the display captures the face of the user for eye-tracking purposes. Unlike specialized 3D displays, our system works with any standard monitor, making it compatible with existing hardware infrastructures in homes, schools, and offices. Moreover, our system also use basic red-cyan filter glasses provide stereoscopic separation. These inexpensive glasses are used to filter the differently colored left and right eye views rendered on the display.

Our system begins by detecting positions of the left eye and right eye of the user by extracting keypoints using the MediaPipe face mesh model~\cite{kartynnik2019real}, which provides 2D coordinates of facial landmarks in real-time. Based on the detected eye positions, the system calculates the viewpoint of the user relative to the display. Then, the system accesses a pre-captured light-field scene, which is a 2D grid of images representing different viewpoints of a scene, and selects the appropriate views that correspond to the current position of each eye. After that, the selected left and right eye views are processed through color channel separation. The left eye view is encoded in the red channel, while the right eye view is encoded in the cyan channels. These are then composited into a single anaglyph image. Finally, the final anaglyph image is displayed on the monitor. As the user moves their head, the tracked eye positions update, and the system seamlessly adjusts the rendered viewpoints to maintain proper parallax effects.

This pipeline creates a responsive system where the 3D visualization adapts to the perspective of the user in real-time, providing motion parallax in addition to the stereoscopic depth cues. This significantly enhances the 3D perception and immersion compared to fixed-viewpoint anaglyph rendering, while maintaining the accessibility and low cost.

\subsection{Eye-tracking Module}
To support dynamic perspective adaptation within the proposed light-field visualization system, a real-time eye-tracking module is integrated as a foundational component. This module is instrumental in synthesizing accurate parallax on a conventional 2D display when viewed through anaglyph glasses. By continuously estimating the user's eyes position, the rendering engine adjusts the synthesized viewpoints to align with the user's line of sight, thereby delivering an immersive and perceptually coherent 3D experience without requiring dedicated stereoscopic hardware.

The eye-tracking functionality is implemented using MediaPipe Face Mesh~\cite{kartynnik2019real}, an advanced computational framework capable of dense, real-time facial landmark detection. MediaPipe identifies 468 three-dimensional facial landmarks, from which a specific subset associated with the periocular region is extracted. The geometric center of each eye is then computed by averaging the coordinates of these landmarks, resulting in a robust estimation of the eye’s central position that is resilient to moderate head movements and orientation changes.

Given the variability in real-time video capture, due to lighting fluctuations, sensor noise, and rapid head motion, a temporal smoothing mechanism is employed to ensure stability and coherence in the estimated eye positions. The system applies a k-frame moving average to the sequence of detected positions, effectively stabilizing the estimation while maintaining responsiveness suitable for interactive rendering. This smoothing strategy achieves a balance between tracking accuracy and latency, ensuring a fluid and consistent user experience.

The stabilized eye position data are forwarded to the rendering module, which dynamically adjusts the light-field rendering to reflect the current viewpoint of the user. This continuous realignment enhances depth perception and spatial accuracy, significantly improving user immersion. The system's ability to track and respond to head and eye movements creates a more natural and realistic 3D effect, even within the limitations of a 2D display environment.

\subsection{Anaglyph Light-field Rendering Module}
\begin{figure}
   \includegraphics[width = \textwidth]{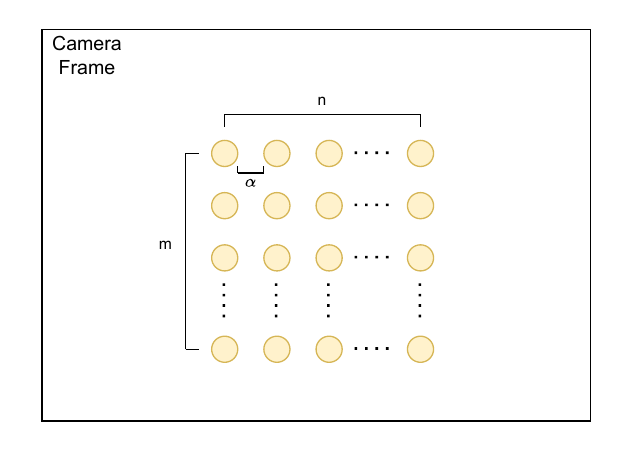}
   \caption{Light-field grid overlays on the camera frame.}
   \label{fig: grid}
\end{figure}

The Anaglyph Light-field Rendering Module is a key part of the proposed 3D visualization system. It creates a 3D effect with depth and parallax by generating stereo images that can be viewed on a regular 2D monitor using anaglyph glasses. Using real-time eye-tracking and anaglyph blending, the system simulates an experience of viewing light-field images in 3D without requiring any specialized display equipment.

As illustrated in Figure~\ref{fig: grid}, light-field data is presented as a two-dimensional $m\times n$ grid, where each point corresponds to an image captured from a unique viewpoint in the scene. The grid is derived from a light-field image which has vertical angular resolution of $m$ and horizontal angular resolution of $n$. This grid structure allows the system to organize multiple perspectives in a simple and efficient way. To establish correspondence between the user’s visual perspective and the light-field image, the system overlays the virtual light-field grid onto the live frame captured by the camera. The coordinates of the user’s left and right pupils are detected in the previous eye-tracking module and mapped to their respective nearest neighbors within the view grid. A configurable parameter $\alpha$ controls the distance between adjacent points in the grid, allowing adjustment of the user's perception about the relative distance of the scene.

Upon determining the optimal pair of views for the left and right eyes, the system synthesizes a single composited anaglyph image. The left eye’s perspective is encoded in the red channel, while the right eye’s view is mapped to the green and blue channels to form the cyan counterpart. When observed through red-cyan anaglyph glasses, this encoding yields a binocular disparity that facilitates the perception of depth. This perceptual effect is a consequence of the brain's fusion of the horizontally offset views into a coherent stereoscopic image.

As head position changes, the eye-tracking module continuously updates the spatial coordinates of each eye. The rendering module dynamically adjusts by re-calculating the nearest-view indices and generating an updated anaglyph image in response. This low-latency feedback loop ensures that the visual output remains consistent with the user’s perspective, delivering a stable and immersive 3D experience.

\section{Experiments}\label{sec: experiment}
\subsection{Implementation Details}
Our proposed system delivers an immersive stereoscopic experience using a standard 2D LCD monitor paired with anaglyph glasses. This setup allows users to perceive depth in light-field imagery without the need for specialized 3D display. A monocular webcam is mounted centrally above the monitor to capture the user's eye movements. This location ensures a clear, symmetrical view of the user's face, improving the accuracy of eyes detection. This camera captures video frames at $30$ frames per second (FPS), providing a continuous stream of data for real-time tracking.

The software stack is developed entirely in Python and leverages two key libraries: Mediapipe and OpenCV. Mediapipe is used for detecting eye landmarks in real-time. OpenCV manages the webcam feed and processes each frame.

All processing tasks, including frame capture, landmark detection, and anaglyph image composition, are performed on the CPU. No GPU acceleration is required, which makes the system compatible with most standard desktop or laptop computers. All the hardware configuration can be found in Table~\ref{tab: hardware_tab}.

\begin{table}
\begin{center}
\caption{Hardware configuration.}
\label{tab: hardware_tab}
\begin{tabular}{|l|l|}
\hline
& Configuration\\
\hline
\textbf{CPU} & Intel(R) Core(TM) i5-9600K CPU @ 3.70GHz\\ 
\textbf{Display} & BenQ GW2765HT, $2560 \times 1440$, 60Hz \\
\textbf{Camera} & Bluelover T3200, $640 \times 480$ \\
\textbf{3D Glasses} & Conventional red-cyan anaglyph glasses \\ 
\hline
\end{tabular}
\end{center}
\end{table}

Moreover, in this experiment, we used light-field images sourced from both synthetic and real-world datasets. Specifically, we used images from the HCI synthetic light-field dataset~\cite{honauer2017dataset} and the EPFL real light-field dataset~\cite{rerabek2016new}. These datasets provide a diverse range of visual content.

\subsection{Rendering Speed}
We evaluated the performance of our 3D visualization system by measuring the frame rate of the entire pipeline and some key components, as shown in Table~\ref{tab: fps_tab}.

The system runs at $30$ FPS, which is the maximum frame rate supported by the integrated camera. Since each frame must be processed as it arrives from the camera, the performance of subsequent modules is tightly coupled to the camera's throughput. However, the processing time for these modules is significantly faster than the FPS of the camera. Therefore, the system can reach the maximum FPS limited by the camera.  To ensure synchronization, all rendering and tracking computations are triggered in response to incoming frames.

The eye-tracking module operates at an average of $1.73$ milliseconds per frame, translating to roughly $578$ FPS. This inference phase was benchmarked on a system without GPU, which provides a reliable baseline for evaluating CPU-bound performance.

Even under these hardware and camera constraints, the system demonstrates consistent low-latency rendering suitable for real-time interaction. Users experience minimal lag during head movement.

\begin{table}
\begin{center}
\caption{FPS of the system and key components.}
\label{tab: fps_tab}
\begin{tabular}{|c|c|c|c|}
\hline
& Eye-tracking Module & Camera & System\\
\hline
FPS & $578$ & $30$ & $30$\\  
\hline
\end{tabular}
\end{center}
\end{table}

\subsection{Grid Distance Affection}
\begin{figure}
   \includegraphics[width = \textwidth]{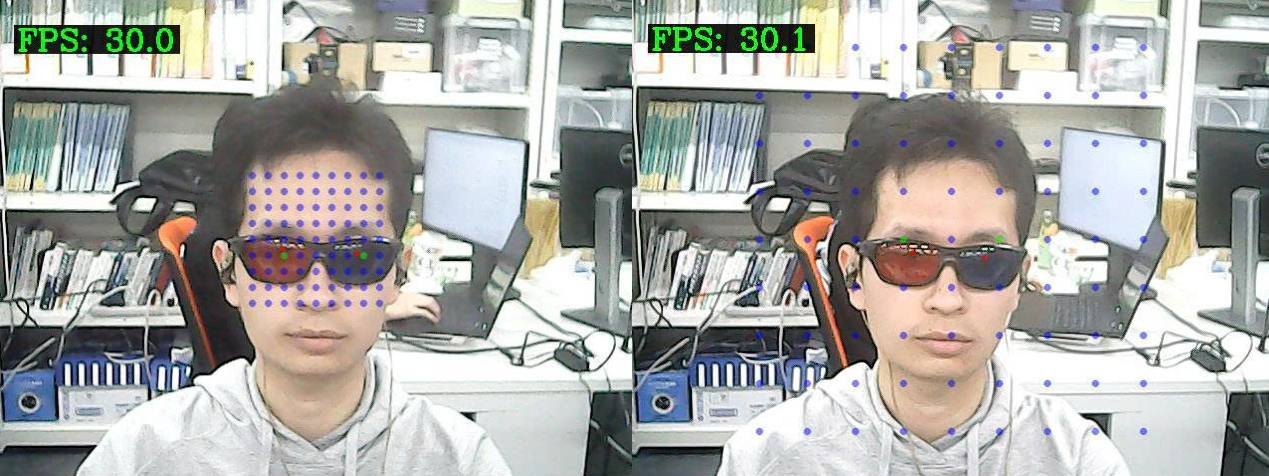}
   \caption{\textbf{Left:} Dense grid configuration with short inter-view distances. \textbf{Right:} Sparse grid configuration with long inter-view distances.}
   \label{fig: dense_sparse}
\end{figure}

The spatial configuration of the light-field views' grid greatly influences both user interaction and viewing experience. We evaluated how different grid spacings impact responsiveness to head movement and the effective viewing zone by comparing two configurations: dense grids with short inter-view distances and sparse grids with larger separations. Figure~\ref{fig: dense_sparse} shows these two types of configurations.

\subsubsection{Dense Grid Configuration}
With minimal spacing, angular views change rapidly in response to subtle head movements. This results in smooth parallax transitions and fine-grained perspective control. However, the trade-off is a limited movement area.

\subsubsection{Sparse Grid Configuration}
Increasing the spacing between views enlarges the effective movement zone, enabling freer head movement and more stable view transitions. However, users may need to move more significantly to notice changes in parallax, which slightly reduces depth responsiveness.

\begin{figure}
   \includegraphics[width = \textwidth]{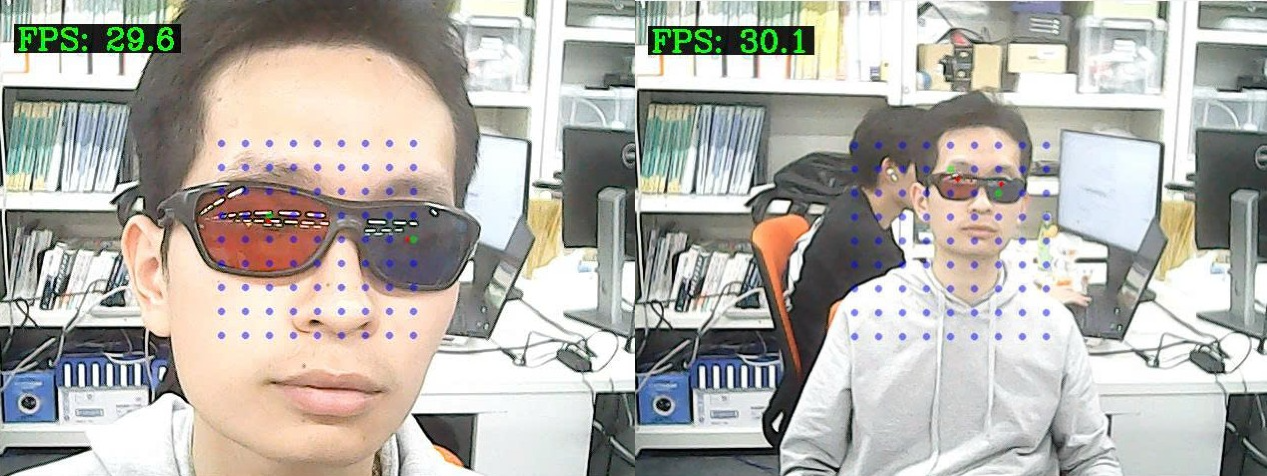}
   \caption{\textbf{Left:} Views mapping when the user is near the monitor. \textbf{Right:} Views mapping when the user is far from the monitor.}
   \label{fig: near_far}
\end{figure}

Viewing distance from the display also affects how the system maps eye positions to views as illustrated in Figure~\ref{fig: near_far}. At close range, the eyes subtend a wider angular separation across the grid, resulting in stronger depth perception. As the user moves farther away, this angular difference shrinks, reducing the stereo baseline and, consequently, the intensity of depth cues. While this can lessen depth vividness, it supports more relaxed, comfortable viewing. This behavior reflects real-world perceptual effects and is consistent with geometric optics and human stereoscopic vision~\cite{allison2009binocular,hibbard2017magnitude}. A moderate grid spacing will offer a practical trade-off between head-tracking responsiveness and user comfort. However, identifying the optimal grid spacing that accurately reflects real-world depth perception remains an open question. Future work will further investigate the relationship between user distance, grid layout, and scene realism to determine how to best configure the system so that the rendered depth matches the actual spatial layout of the scene.

\section{Discussion}\label{sec: discussion}
This work demonstrates that real-time 3D light-field viewing can be achieved using widely available, low-cost hardware. Our system leverages a basic RGB webcam, a conventional display, and runs entirely on CPU, making it accessible to users without specialized setups. By maintaining a consistent frame rate of $30$ FPS, the system offers a smooth, interactive experience that enhances depth perception and user immersion.

Nevertheless, several limitations need to be addressed. Anaglyph glasses, while affordable and easy to use, reduce overall visual quality by distorting colors and introducing ghosting artifacts. These limitations may reduce the realism of the 3D effect and hinder adoption for applications requiring accurate color representation. Moreover, determining the optimal grid spacing for light-field images is also a challenge. The grid separation parameter directly affects how accurately the system can replicate real-world depth perception. Furthermore, even if an optimal grid spacing is chosen, the constrained movement within the grid can still detract from the user's experience. Users may feel restricted in their movement, leading to a less immersive interaction. Another major challenge is the limited angular resolution of most available light-field datasets. These datasets often fail to provide the fine granularity necessary for seamless motion parallax. As a result, users may perceive small jitter jumps in perspective when moving their heads, which can reduce immersion and even cause discomfort over extended periods.

Despite these challenges, the system holds significant promise for practical use. In educational settings, it could allow students to explore complex spatial concepts in a more intuitive and engaging way. In digital media and the arts, it offers creators a platform to produce and share immersive 3D content without high equipment costs. Additionally, for accessibility-focused applications, it enables users with limited access to VR/AR hardware to experience depth-enhanced visualizations using their existing devices.

A possible direction for future research is enhancing angular density of light-field images to reduce visual artifacts from sparse scenes and allow smoother transitions. It will enable the user to experience a larger freedom of head movements. Additionally, exploring compatibility with real-time light-field video capture systems could enable live, dynamic content applications.

\section{Conclusion}\label{sec: conclusion}
This paper proposed a low-cost 3D visualization system for light-field images, running smoothly at $30$ FPS. It significantly lowers the barrier to experiencing immersive 3D content by removing the need for specialized displays or VR headsets. It provides a practical solution for accessible, low-cost 3D visualization and paves the way for future advancements in user-friendly 3D interaction systems built on everyday hardware.

%
%
%
\bibliographystyle{splncs04}
\bibliography{mybibliography}

\end{document}